\documentclass[11pt]{article}
\usepackage{moriond,epsfig}

\def\Journal#1#2#3#4{{#1} {\bf #2}, #3 (#4)}

\def\PRL{\em Phys. Rev. Lett.}
\def\PRD{{\em Phys. Rev.} D}

\def\D0{D\O}

\def\be{\begin{equation}}
\def\ee{\end{equation}}
\def\bea{\begin{eqnarray}}
\def\eea{\end{eqnarray}}

\begin{document}
\vspace*{4cm}
\title{TOP QUARK PAIR PRODUCTION AT THE TEVATRON
}
\author{Jason Nielsen \\ (on behalf of the CDF and \D0 collaborations)}

\address{Physics Division \\ Lawrence Berkeley National Laboratory \\ 
One Cyclotron Road, Berkeley, CA  94720 U.S.A.}

\maketitle\abstracts{
The measurement of the top quark pair production cross section in
proton-antiproton collisions at $\sqrt{s}=1.96$ TeV is a test of
quantum chromodynamics and could potentially be sensitive to new
physics beyond the standard model.  I report on the latest $t\bar t$
cross section results from the CDF and \D0 experiments in various final
state topologies which arise from decays of top quark pairs.
}

\section{Introduction to top quark production and decay}

The physics of the top quark, the heaviest known fundamental particle,
offers a new testing ground for the standard model,
including quantum chromodynamics (QCD), as well as a new frontier for
unexpected physics beyond the standard model.  The top quark
production measurements by the CDF and \D0 experiments on the Tevatron
collider at Fermilab are the first step towards exploiting this
promising physics sector.

Top quark pair production in $p\bar p$ collisions at
$\sqrt{s}=1.96\,\mathrm{TeV}$ proceeds predominantly through the
quark-antiquark annihilation (85\%) and gluon fusion (15\%) diagrams.
The most recent QCD calculations of the cross section yield 
$6.7^{+0.9}_{-0.7}\,\mathrm{pb}$ for a
$175\,\mathrm{GeV}/c^2$ top quark mass~\cite{bib:cacciari,bib:kidonakis}.
Each top quark decays immediately to a $W$ boson and $b$ quark, and
the decays of the $W$ bosons to quarks or leptons define the event
signature of the top quark in the following three different final
states: dilepton, with 2 leptons and 2 jets; lepton plus jets,
with 1 lepton and 4 jets; and all-jets, 
with 0 leptons and 6 jets.  In each final state, 2 of the jets are $b$
jets which can be tagged using heavy-flavor tagging algorithms based
on identifying semileptonic $b$ decays or tracks from $b$ decays which
do not originate from the primary event vertex.

\section{Results from CDF and \D0 experiments at the Tevatron}

Both the CDF and \D0 collaborations are pursuing measurements in all 3
final states using a variety of techniques.  One challenge inherent in
all measurements is the determination of the data sample composition.
Knowledge of the sample composition is also important for future
studies of top quark properties.  The results presented here use
collected data which represent integrated luminosities of
$150-200\,\mathrm{pb}^{-1}$.   All of the results estimate the number
of $t\bar t$ events in the data sample and use the $t\bar t$ signal
acceptance, determined using detailed detector simulation, to
calculate a cross section.

\subsection{Final states with two leptons plus jets}

The dilepton sample is a small (5\% branching fraction for $e/\mu$
analyses) but clean sample with backgrounds from Drell-Yan production
and fake leptons.  Analyses focus either on tight requirements to
select an extremely clean sample ($b$-tagged $e+\mu$ channel) or
loosened lepton identification (lepton + isolated track channel) in
order to reduce statistical uncertainty~\cite{bib:cdfdil}.

\begin{figure}
\begin{center}
\epsfig{figure=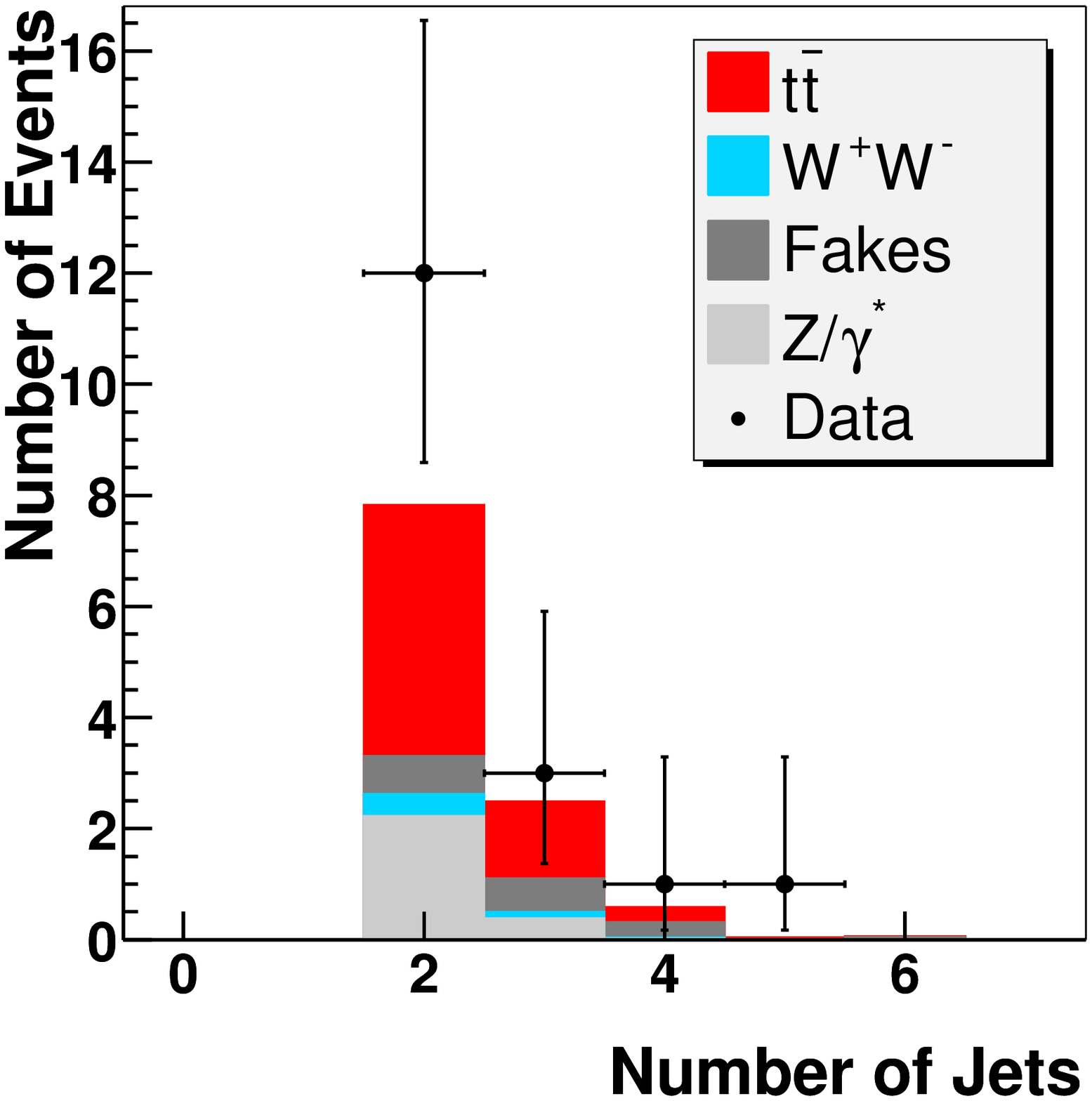,width=0.48\textwidth}
\epsfig{figure=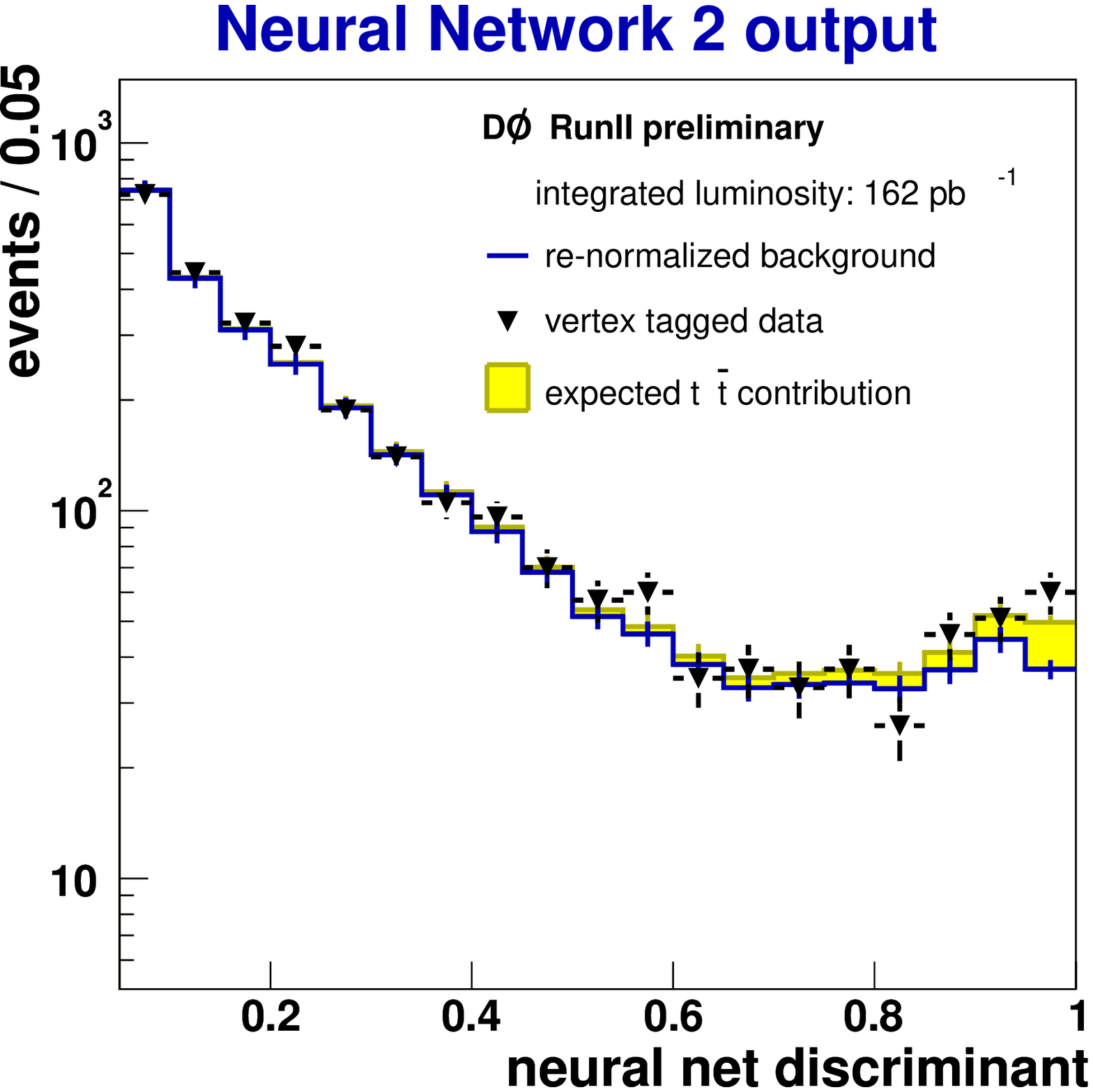,width=0.48\textwidth}

\epsfig{figure=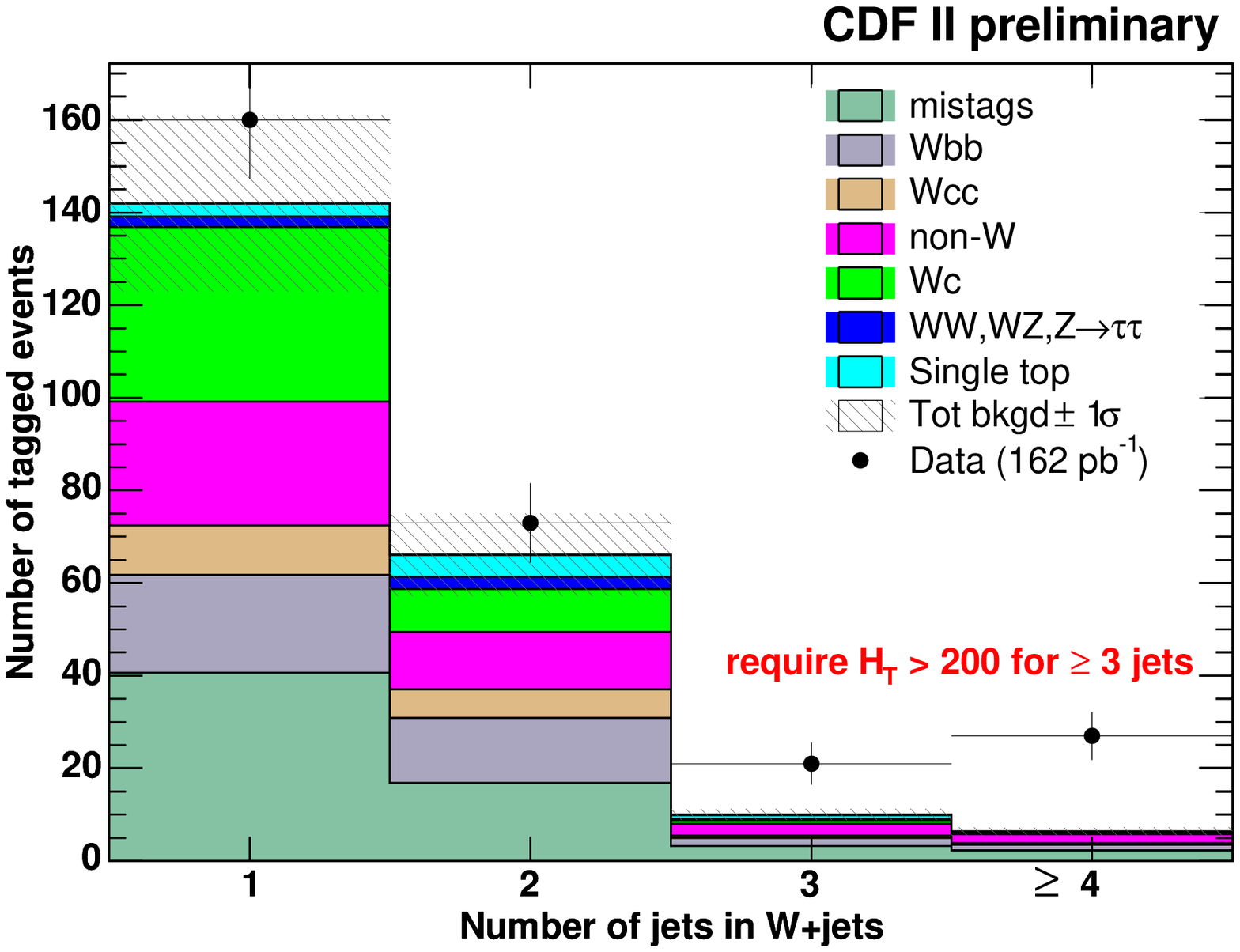,width=0.48\textwidth}
\epsfig{figure=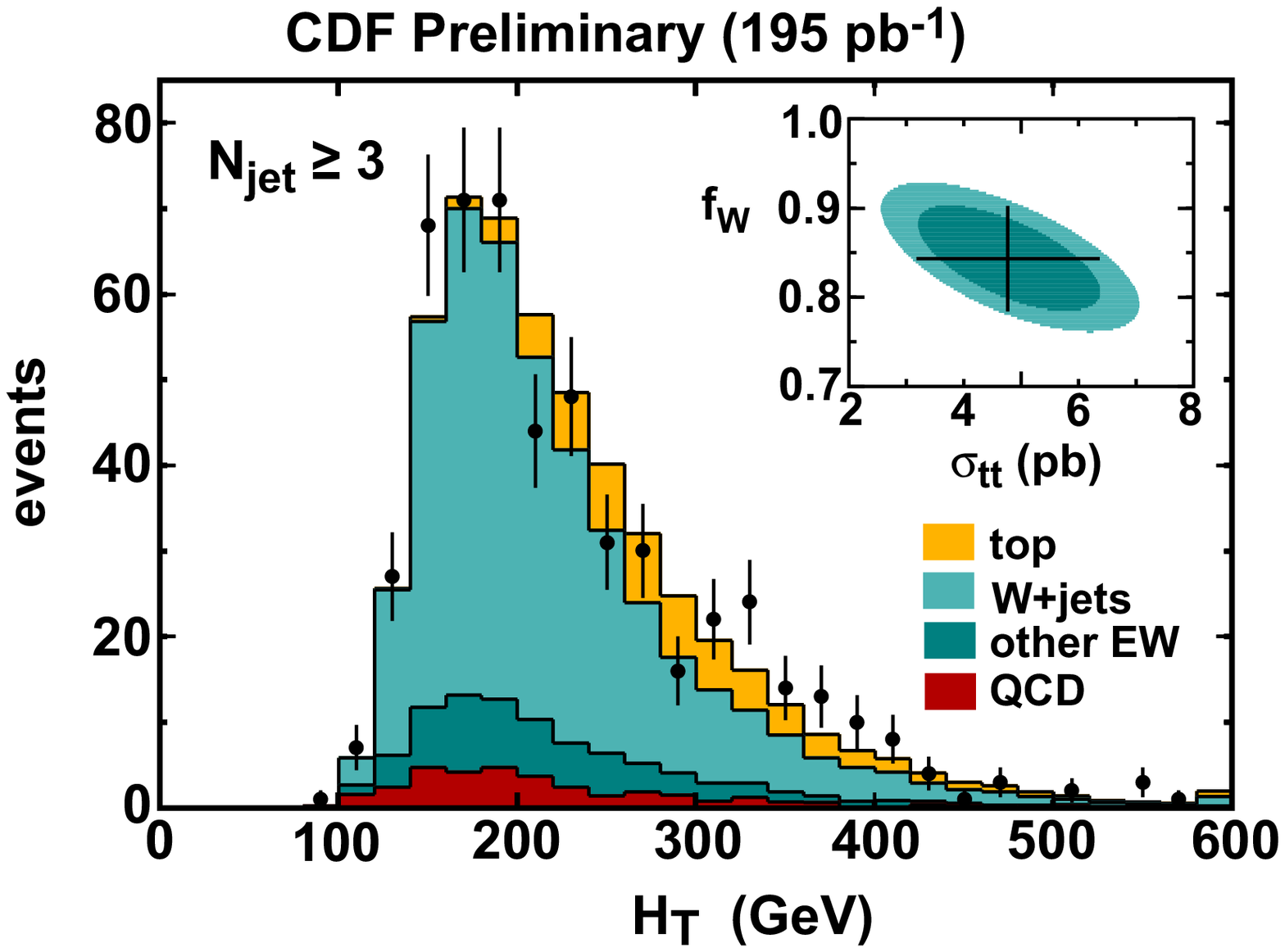,width=0.48\textwidth}
\caption{Top pair production cross section results from \D0 dilepton
  jet-counting analysis (top left), \D0 all-jets artificial neural
  network analysis (top right), CDF jet-counting lepton plus jets
  analysis (bottom left), and CDF $H_T$-fitting lepton plus jets
  analysis (bottom right).
\label{fig:analyses}}
\end{center}
\end{figure}

\subsection{Final states with one lepton plus jets}

The lepton + jets signature benefits from high lepton trigger
efficiency combined with large branching fraction (30\% for $e/\mu$
plus jets).  The dominant non-top backgrounds in the sample are
$W$+jets production and non-W QCD events, but these backgrounds
can be estimated by fitting kinematic distributions, e.g., the total
transverse energy ($H_T$) in selected events~\cite{bib:d0ljetskine}.
Since the $t\bar t$ signal is expected to have 4 jets, a sample of
events with lepton plus 2 jets is used as a control region for
background checks.

The $W$+jets and non-$W$ QCD backgrounds can be reduced by requiring
at least one jet in the event be $b$-tagged, as expected from
$t\rightarrow Wb$ decay.  Even though the $b$-tagging requirement is
only 50\% efficient for $t\bar t$ events, approximately 98\% of
$W$+light flavor events are rejected.  The remaining contribution
due to $W$+heavy flavor production is estimated by calculating the
heavy flavor fraction in $W$+jets events and applying this fraction to
the observed data sample~\cite{bib:cdflj,bib:d0ljetsbtag}.

\subsection{All-jet final state}

The all-jets signature presents a challenge for triggering; one
approach is to trigger on the total event energy and the number of
high-$E_T$ jets.  Because of the large branching fraction (44\% of
$t\bar t$ events do not have leptonic $W$ decays), this channel has the
largest number of $t\bar t$ events.

The overwhelming background of QCD jet production is estimated in one
of two ways.  An artificial neural network can be employed to
discriminate $t\bar t$ events from background based on kinematic
variables which are most sensitive to high-mass objects like top
quarks.  These include total transverse energy, dijet masses and top
quark masses.  Alternatively, a tag rate matrix can be constructed
from the control sample of 4-jet events (where there is no signal
expected) and applied to the 6-jet sample.

\section{Summary of pair production cross section results\label{sec:summary}}

\begin{figure}[htbp]
\begin{center}
\begin{minipage}{0.48\textwidth}
\epsfig{figure= 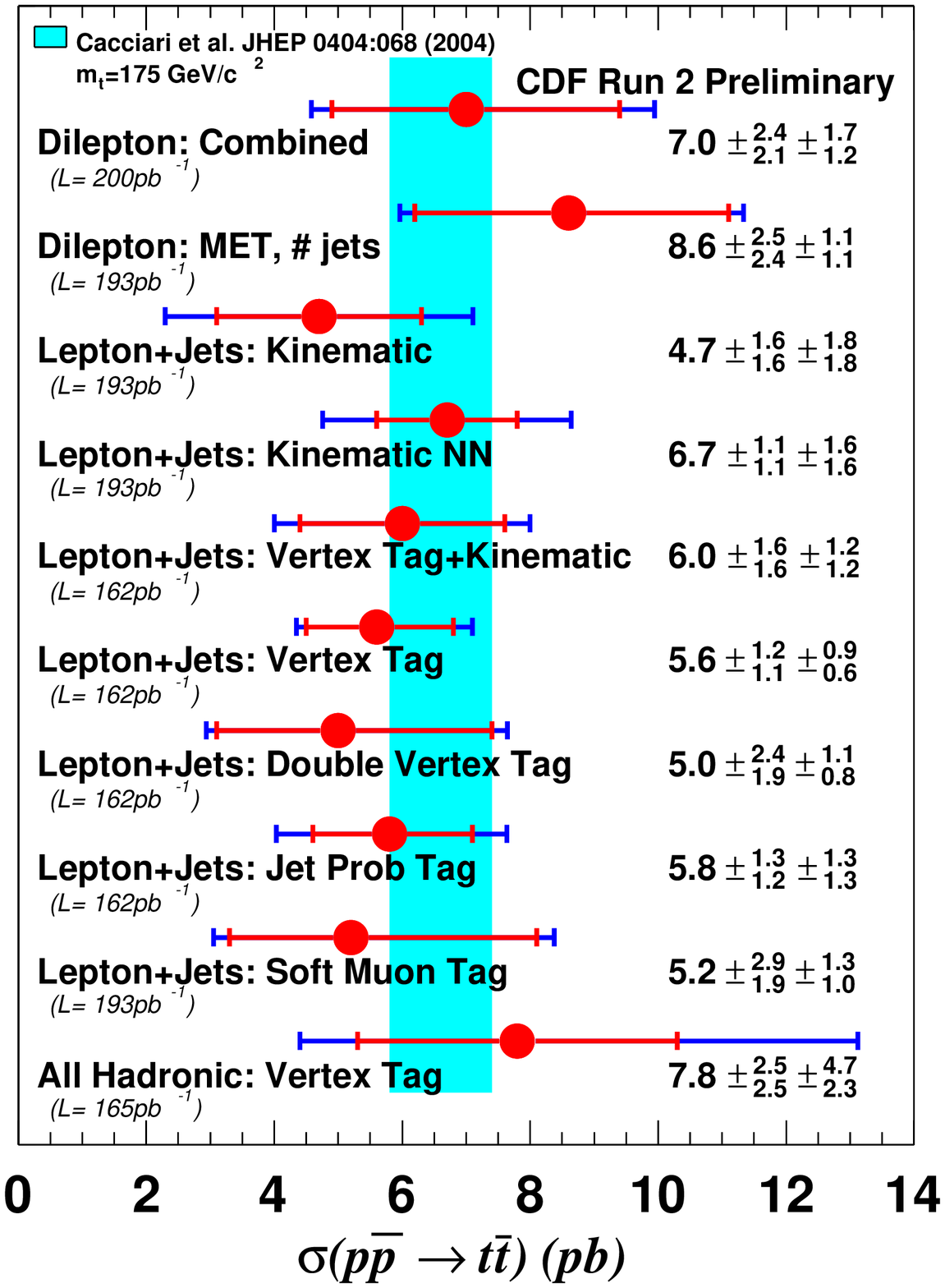,width=\textwidth}
\end{minipage}
\begin{minipage}{0.48\textwidth}
\epsfig{figure= 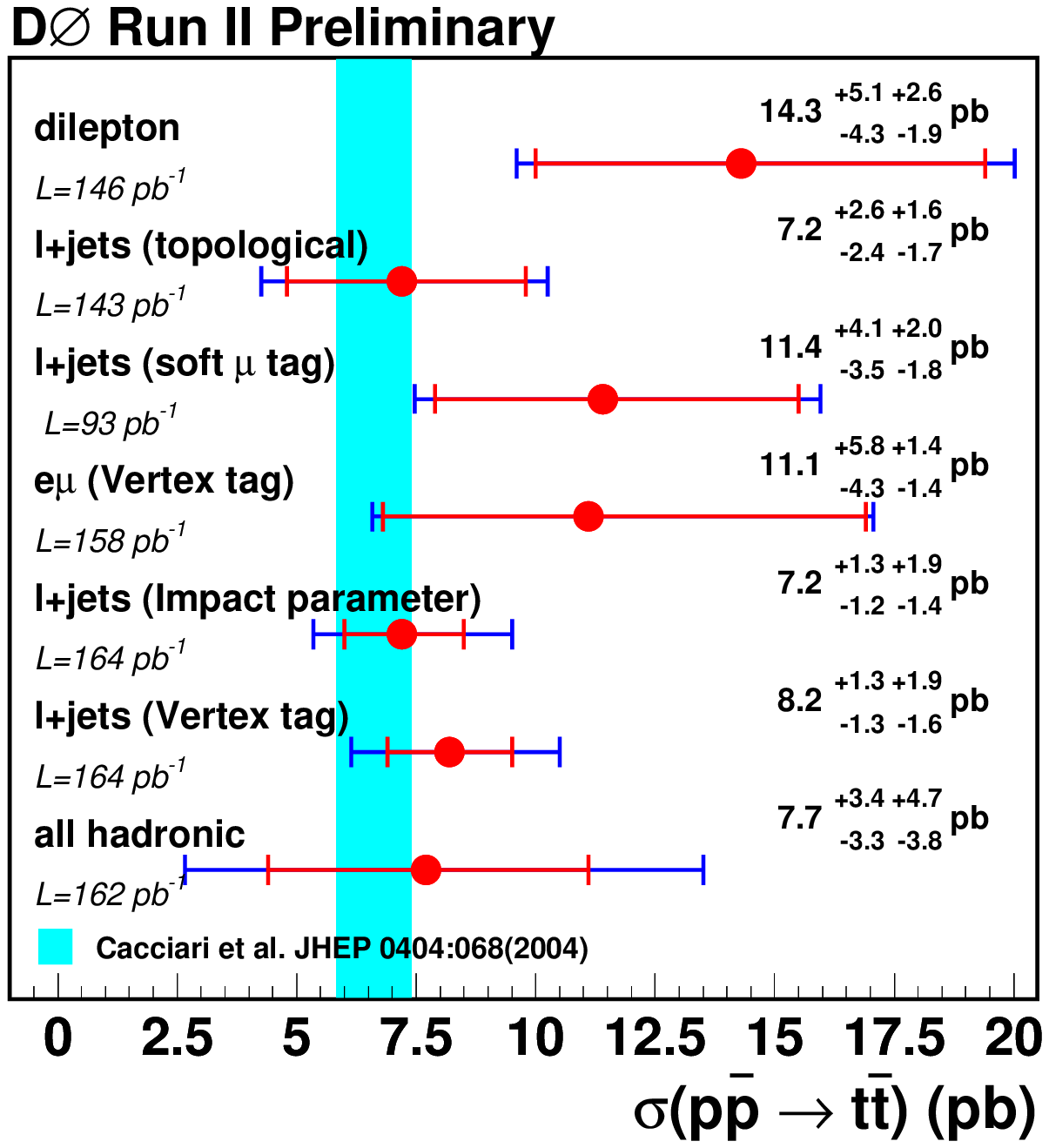,width=\textwidth}
\end{minipage}
\caption{Summary of top pair production experimental results from CDF
  (left) and \D0 (right) experiments.  The blue band on both plots represents the
  NLO QCD theory calculation for a $175\,\mathrm{GeV}/c^2$ top mass.
\label{fig:allresults}}
\end{center}
\end{figure}

\begin{figure}[htbp]
\begin{center}
\epsfig{figure=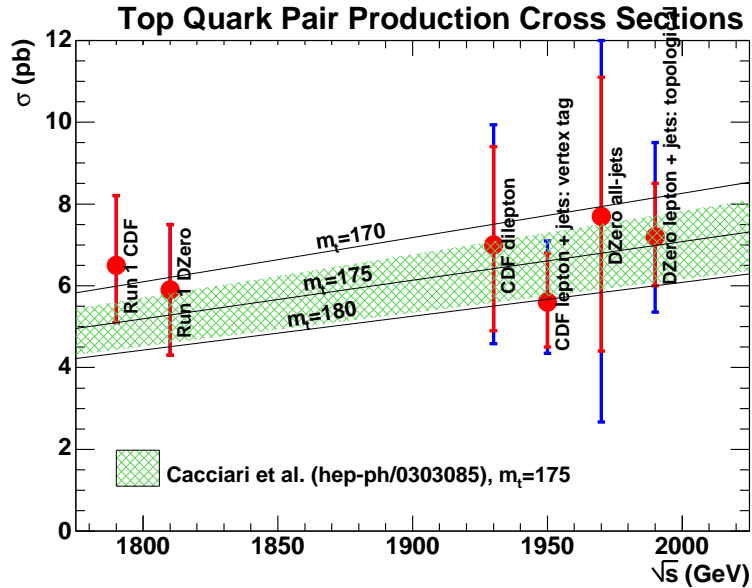,width=0.7\textwidth}
\caption{Comparison of selected results with NLO QCD
  calculations as a function of center-of-mass energy.
\label{fig:theorycomparison}}
\end{center}
\end{figure}

The summary of all results from CDF and \D0 are displayed in
Fig.~\ref{fig:allresults}.  All of the measurements are consistent
with each other and with the theory calculation, at least within the
quoted errors.  
Selected experimental results at $\sqrt{s}=1.8\,\mathrm{GeV}$ and
$1.96\,\mathrm{GeV}$ are compared to the theory calculation at those two
center-of-mass energies in Fig.~\ref{fig:theorycomparison}.
It is reasonable to expect that additional collected data and further
understanding of the sample composition will reduce the uncertainties
on all of these measurements.

\section*{Acknowledgments}

It is a pleasure to thank the organizers for fostering the stimulating
atmosphere of the conference.  I must also acknowledge my
CDF and \D0 colleagues for their help in preparing these results.


\section*{References}
\begin{thebibliography}{99}

\bibitem{bib:cacciari} M. Cacciari {\it et al.}, \Journal{JHEP}{404}{68}{2004}.

\bibitem{bib:kidonakis} N. Kidonakis and R. Vogt, \Journal{\PRD}{68}{114014}{2003}.

\bibitem{bib:cdfdil} D. Acosta {\it et al.} (CDF collaboration),
\Journal{\PRL}{93}{142001}{2004}.

\bibitem{bib:d0ljetskine} V.M. Abazov {\it et al.} (\D0
collaboration), ``Measurement of $t\bar t$ cross section using l+jets
kinematic characteristics,'' {\tt hep-ex/0504043},
submitted to Phys. Rev. Lett.

\bibitem{bib:cdflj} D. Acosta {\it et al.} (CDF collaboration),
\Journal{\PRD}{71}{052003}{2005}.

\bibitem{bib:d0ljetsbtag} V.M. Abazov {\it et al.} (\D0
collaboration), ``Measurement of $t\bar t$ cross section using l+jets
events with lifetime b-tagging,'' {\tt hep-ex/0504058},
submitted to Phys. Rev. Lett.

\end{thebibliography}
\end{document}